\documentstyle[12pt,epsfig]{article}
\begin{document}

\title
{Contribution of boundness and motion of nucleons to the EMC effect}
\author{
 B.L. Birbrair, M.G. Ryskin and V.I. Ryazanov\\
 Petersburg Nuclear Physics Institute\\
Gatchina, St. Petersburg 188300, Russia}
\date{}
\maketitle

\begin{abstract}

The kinematical corrections to the structure function of nucleon in
nucleus due to the boundness and motion of nucleons arise from
the
excitation of the doorway states for one-nucleon transfer reactions
in
the deep inelastic scattering on nuclei.

\end{abstract}

\section{Introduction}

It is known more than 20 years that the cross section of deep
inelastic
scattering (DIS) on nuclear target is not equal to the sum of  cross
sections on free nucleons \cite{EMC}. This means that the
interaction
inside the nucleus distorts the parton distribution in a nucleon.
 But at first it is necessary to single out the kinematical effects arising
 from the boundness and motion of nucleons in nuclei because otherwise it is
 hardly possible to conclude what actually happends with the nucleon structure
 functions in nuclear matter. This is the aim of our work rather than the
 explanation of the EMC effect. The above kinematical effect is due to
  the fact
that the four-momentum of nucleon in nucleus is not equal to that
of a free nucleon. Indeed,
 the heavy photon ($\gamma^*$) is absorbed by
a single nucleon and the Deep Inelastic Scattering (DIS) proceeds via
the following stage:
$$l\ +\ A\rightarrow l'\ +\ X\ +\ (A-1)^*\; ,$$
where $l$ and $l'$ denote the incoming and outgoing leptons, $X$ - the
final hadronic state after the $\gamma^*$-nucleon interaction and $A$
 is the target nucleus. Before absorbing the heavy photon
$(\gamma^*$) the struck nucleon has a certain energy-momentum
distribution in nucleus. Besides this the ``residual'' ($A-1$) nucleus
is excited.\\

There were a few attempts to account for the Fermi motion,
boundness
and the change of the effective $\gamma$-nucleon flux factor inside
the
nucleus \cite{BR,Vagr,FS} (see \cite{Arn} and references therein for
more details). They all were based on a seemingly obvious
assumption
that the energy-momentum distribution of the struck nucleon is
described by the ground-state one-nucleon spectral function.
\begin{equation}
S_g({\bf p},\varepsilon)=\langle A_0|a^+({\bf p})\delta\bigg(
\varepsilon+H-{\cal E}_0(A)\bigg)a({\bf p})|A_0\rangle\ ,
\end{equation}
where $|A_0\rangle$ is the ground state of target nucleus $A$,
$a({\bf
p} )$ and $a^+({\bf p})$ are operators of nucleon with the momentum
${\bf p}$ (the spin and isospin variables are omitted), $H$ is the
nuclear Hamiltonian in the second quantization and
${\cal E}_0(A)$
is the ground-state binding energy of nucleus $A$. The calculations
\cite{Vagr} were performed by using the following semiempirical
model
for the quantity (1): the nucleon energy distributions were described
by the experimental data on the separation energies of protons
from the
$(e,e'p)$ reactions \cite{6} (the difference between the proton and
neutron separation energies was neglected leading to about 10\%
error)
and calculating the momentum distributions within the harmonic
oscillator model with the parameter
$\hbar\omega_0=(45\,A^{-1/3}-25\,A^{-2/3})$ MeV reproducing the
observed rms
radii of nuclei.

However nobody realized in this connection that the DIS on nuclei is
rapid process, and therefore the energy-momentum distribution of
struck
nucleon is described by the spectral function of nuclear mode
which is
excited via a sudden perturbation rather than that of the ground
state.
Our work is based on the fact that the relevant mode is provided by
the
doorway states for the one-nucleon transfer reactions. As
demonstrated
in \cite{3}--\cite{BR3} the above states are eigenstates of
nucleon
in the static nuclear field.

Recall that the microscopic nuclear models are based on certain
approximations for the in-medium nucleon mass operator $M$. For
instance the nuclear shell model potential is the approximation for
the
mass operator at nuclear Fermi-surface, the optical model potential
is
dealing with the mass operator at low and intermediate energies,
etc.
In all the available approaches
the mass operator includes all the Feynman
diagrams which
are irreducible in the one-particle channel, and therefore it cannot be
calculated. Instead it is described by a set of phenomenological
parameters.

In contrast to the above models the nuclear static potential is the
mass operator at the infinite value of the energy variable. Only the
Hartree diagrams with the free space ({\em i.e.} vacuum) nuclear
forces
survive in this case thus permitting the model-independent
calculation
of the static field. So the doorway states (DS) under consideration
appear to be the unique nuclear object, both model-independent and
described by the exactly soluble problem.

The calculation \cite{BR2} showed that the rms radii of the DS
density
distributions are appreciably less than those of the ground-state
ones:
for instance, the  value of $\langle
r^2\rangle=A^{-1}\int\rho(r)r^2d^3r$ is $\rm10.88\,fm^2$ for the
ground
state of $^{40}$Ca being only $\rm8.76\,fm^2$ for the DS. As a
result
the nucleon motion ({\em i.e.} the value of $\langle{\bf p}^2\rangle$)
was underestimated in \cite{Vagr} by about 25\% \footnote{By the same
reason the kinematical effect was underestimated in Ref.\cite{BLev} as
well.}.\\

In Sec. 2 we briefly describe the formalism of the DS. In Sec. 3 the
DIS structure functions $F_2$ are calculated for different nuclei and
deuteron; to single out the boundness and motion effects we
disregarded
the possible changes of the parton distribution inside the nucleon in
nucleus. In the last Sec. 4 we compare the calculated ratios
$2F_{2A}/AF_{2D}$ with the available experiments.


\section{Doorway states for one-nucleon transfer reactions}

\subsection{Theory}

Evolution of the state arising from the one-nucleon transfer to the
nuclear ground state $|A_0\rangle$ at the initial time moment $t=0$ is
described by the single-particle propagator \cite{7}
\begin{eqnarray}
&& S(x,x';\tau)\ =\ -i\langle A_0|T\psi(x,\tau)\psi^\dag(x',0)|A_0
\rangle\ = \nonumber\\
= && i\theta(-\tau)\sum^{(A-1)}_j
\Psi_j(x)\Psi^\dag_j(x')e^{-iE_j\tau}-i\theta(\tau) \sum^{(A+1)}_k
\Psi_k(x)\Psi^\dag_k(x')e^{-iE_k\tau}.
\end{eqnarray}
At $\tau<0$ it describes the evolution of the hole state,
\begin{equation}
\Psi_j(x)=\langle(A-1)_j|\psi(x)|A_0\rangle\ ,\quad E_j=
{\cal E}_0(A)-{\cal E}_j(A-1)\ ,
\end{equation}
when the nucleon is removed from the ground state $A_0$, whereas at
$\tau>0$ the evolution of the particle state is described,
\begin{equation}
\Psi_k(x)=\langle A_0|\psi(x)|(A+1)_k\rangle\ , \quad E_k=
{\cal E}_k(A+1)-{\cal E}_0(A)\ ,
\end{equation}
when nucleon is added to the ground state $A_0$. The quantities
${\cal E}_j(A-1)$, ${\cal E}_k(A+1)$ and ${\cal E}_0(A)$ are total
binding energies of the states $|(A-1)_j\rangle$ of the $(A-1)$
nucleus, the states $|(A+1)_k\rangle$ of the $(A+1)$ nucleus and the
ground state $|A_0\rangle$ of the $A$ one.

The Fourier transform of the propagator
\begin{equation}
G(x,x';\varepsilon)=\int S(x,x';\tau)e^{i\varepsilon\tau} d\tau =
\sum^{(A-1)}_j\frac{\Psi_j(x)\Psi^\dag_j(x')}{\varepsilon-E_j-i\delta}
+\sum^{(A+1)}_k\frac{\Psi_k(x)\Psi^\dag_k(x')}{\varepsilon-E_k+i\delta}\
,
\end{equation}
$\quad\hspace{10cm} \delta\to\ +0$\\
obeys the Dyson equation
\begin{equation}
\varepsilon G(x,x';\varepsilon)\ =\ \delta(x-x')+\hat k_xG(x,x';
\varepsilon)+ \int M(x,x_1;\varepsilon)G(x_1,x';\varepsilon)dx_1\ ,
\end{equation}
where $\hat k_x$ is the kinetic energy and the mass operator $M(x,x';
\varepsilon)$ includes all Feynman diagrams which are  irreducible in
the one-particle channel.

We are interested in the very beginning of the evolution, i.e. the
$\tau\to0$ limit. According to the time-energy Heisenberg relation this
is equivalent to the limit $\varepsilon\to\infty$. In this limit
\begin{equation}
G(x,x';\varepsilon)\ =\ \frac{I_0(x,x')}\varepsilon +\frac{I_1(x,x')}{
\varepsilon^2}+\frac{I_2(x,x')}{\varepsilon^3}+\cdots\ ,
\end{equation}
where (see the definition (1) of the propagator)
\begin{equation}
I_0(x,x')\ =\ \sum^{(A-1)}_j\Psi_j(x)\Psi_j^\dag(x')+\sum^{(A+1)}_k
\Psi_k(x)\Psi^\dag_k(x')=i\bigg[S(x,x';+0)-S(x,x';-0)\bigg];
\end{equation}
\begin{eqnarray}
I_1(x,x')&&=\ \sum^{(A-1)}_jE_j\Psi_j(x)\Psi_j^\dag(x')+\sum^{(A+1)}_k
E_k\Psi_k(x)\Psi^\dag_k(x')\ =\nonumber\\
&&=\ -\bigg[\dot
S(x,x';+0)-\dot S(x,x';-0)\bigg];
\end{eqnarray}
\begin{eqnarray}
I_2(x,x')&&=\
\sum^{(A-1)}_jE^2_j\Psi_j(x)\Psi_j^\dag(x')+\sum^{(A+1)}_k
E^2_k\Psi_k(x)\Psi^\dag_k(x')\ =\nonumber\\
&&=\ -i\bigg[\ddot S(x,x';+0)-\ddot
S(x,x';-0)\bigg]
\end{eqnarray}
the quantities $I_0, I_1$ and $I_2$ thus describing the very beginning
of the evolution $\left(\dot S=\frac{\partial S}{\partial\tau}, \ddot S
=\frac{\partial^2S}{\partial\tau^2}\right)$.

Now consider the mass operator $M(x,x';\varepsilon)$. It includes the
energy-independent Hartree diagrams $U_{st}(x)\delta(x-x')$ (which were
shown in Fig.3 of Ref.\cite{BR3}) the higher-order diagrams describing
the nuclear correlation effects (the lowest-order diagram of such kind
was shown in Fig.4a of Ref.\cite{BR3}) and the Fock ones (Fig. 4b of
Ref.\cite{BR3}).  The correlation diagrams include the propagators of
intermediate states thus behaving as $\varepsilon^{-1}$ in the
$\varepsilon\to\infty$ limit (see Ref.\cite{8} for more stringent
demonstration). The same is valid for the Fock diagrams.
Indeed, the interaction between baryons proceeds via the exchange by
some particles (they are quark--antiquark pairs and/or gluons in the
QCD) and therefore both the momentum and the energy are transferred
through the interaction. As a result the Fock diagrams also include the
intermediate state propagators thus being of order of
$\varepsilon^{-1}$ in the $\varepsilon\to\infty$ limit. (In Ref.
\cite{3} this is demonstrated for the meson-nucleon intermediate
state). So the mass operator in this limit is \begin{eqnarray} &&
M(x,x';\varepsilon)\ =\ U_{st}(x)\delta(x-x')+ \frac{\Pi(x,x')
}\varepsilon\ +\ \cdots \\ &&\ \varepsilon\to\infty \nonumber
\end{eqnarray} Introducing the static Hamiltonian \begin{equation}
h_{st}\ =\ \hat k_x+U_{st}(x) \end{equation} let us write down the
high-energy limit Dyson equation in the form \begin{equation}
\varepsilon G(x,x';\varepsilon)=\delta(x-x')+h_{st}G(x,x';\varepsilon)
+\int\left(\frac{\Pi(x,x_1)}\varepsilon+\cdots\right)
G(x_1,x';\varepsilon)dx_1\ .
\end{equation}
Putting into (13) the asymptotics (7) and equating coefficients at the
same powers of $\varepsilon^{-1}$ we get
\begin{eqnarray}
&& \sum^{(A-1)}_j\Psi_j(x)\Psi^\dag_j(x')+\sum^{(A+1)}_k\Psi_k(x)
\Psi^\dag_k(x')\ =\ \delta(x-x')\\
&& \sum^{(A-1)}_jE_j\Psi_j(x)\Psi^\dag_j(x')+\sum^{(A+1)}_k E_k
\Psi_k(x)\Psi^\dag_k(x')\ =\ h_{st}\delta(x-x')
\end{eqnarray}
\begin{equation}
 \sum^{(A-1)}_jE^2_j\Psi_j(x)\Psi^\dag_j(x')+\sum^{(A+1)}_k E^2_k
\Psi_k(x)\Psi^\dag_k(x')\ =\ h^2_{st}\delta(x-x')+\Pi(x,x').
 \end{equation}
Equations (9), (12) and (15) may be written as
\begin{equation}
-\left[\dot S(x,x';+0)-\dot S(x,x';-0)\right]\ =\ h_{st}\delta(x-x')\
=\ \left[k_x^{}+U_{st}(x)\right]\delta(x-x')\ .
\end{equation}
As follows from the lhs of (17) the hamiltonian $h_{st}$ describes the
very beginning of the one-nucleon transfer process the eigenstates of
$h_{st}$ thus being the doorway states for one-nucleon transfer
reactions. On the other hand the rhs of (17) shows that the hamiltonian
$h_{st}$ describes  the motion of nucleon in nuclear static field
$U_{st}(x)$. Indeed, the latter is expressed through the free-space
$NN$ forces rather than the effective ones thus being the nucleon field
rather than the quasiparticle one. So we proved that the doorway states
for one-nucleon transfer fast reactions
are the
eigenstates of nucleon in nuclear static field.

\subsection{Doorway eigenfunctions}
Since the doorway states (DS) describe the motion of the nucleon in
nuclear {\it static} field the corresponding eigenfunctions may be
calculated in a model independent way. Indeed, the two-particle forces
are determined from the experimental data on the elastic
nucleon-nucleon scattering
(i.e. from the phase shifts analysis)\cite{6b} and the deuteron
properties. The necessary information about the multiparticle forces is
obtained from the observed energy spectra of the doorway states
\cite{3}. So the only additional information needed for
calculation of the static field in a given nucleus is that on the
nucleon density distributions in this nucleus. In all the nuclei which
are treated in the present paper these distributions are spherically
symmetric thus leading to the static field with the same symmetry.
Hence the quantum mechanical problem is the motion of a particle in a
central field. This problem is solved with any desired accuracy and
without any simplifications.\\

 We have to emphasize that the doorway states are not
the eigenfunctions of the
total nuclear Hamiltonian thus being
fragmented over the actual nuclear states owing to the
 correlation effects. The observed spreading width of the DS is
about 20 MeV; that is the relaxation time $\sim 0.3\cdot 10^{-22}$
sec. This is much larger than the time chracteristic for DIS
which is of the order of $2q_0/Q^2 \simeq 1/mx \sim 3\cdot 10^{-24}$
sec in the nucleus rest frame.  So during the DIS process the DS do
not have time to be distorted by the correlations thus permitting the
exact account for the nucleon boundness and motion to the EMC effect.\\

The relevant energy-momentum distribution of nucleons for DIS is
determined by the spectral function of the DS (rather than the ground
state one):
\begin{equation}
S_{DS}(\varepsilon,\vec p)\; =\; S_{p}(\varepsilon,\vec p)\; +\;
S_{n}(\varepsilon,\vec p)\; ,
\end{equation}
where the proton spectral function is
\begin{equation}
S_{p}(\varepsilon,\vec p)\; =\;\frac 1{4\pi}\sum^{(p)}_\lambda
\nu_\lambda f_\lambda(\vec p)\delta(\varepsilon-\varepsilon_\lambda)\;
.
 \end{equation}
The sum in the r.h.s. runs over the proton DS,
$\lambda$ stands for the angular momentum $j$ and other quantum numbers
of a particle state in central field, $\nu_\lambda$ equal to $2j+1$ for
the filled states and the actual number of nucleons on partly filled
ones, $\varepsilon_\lambda$ are the DS energies and $f_\lambda(p)$ are
found by solving the Dirac equation (see Ref.s\cite{3,BR3} for
details).  \begin{equation} h_{st}\psi_\lambda(\vec
r)\;=\;\varepsilon_\lambda\psi_\lambda(\vec r)\; .  \end{equation} The
function $f_\lambda(p)=u^2_\lambda(p)+w^2_\lambda(p)$, given by the sum
of the upper and lower components square of the bi-spinor
$\psi_\lambda(p)$ (in momentum space), is normalized by the condition
\begin{equation} \int f_\lambda(p)p^2dp=1\; .  \end{equation}

The neutron spectral function obeys the same relation in which the
proton DS are substituted by the neutron ones.\\

It is instructive to mention that the spectral functions
$S_{DS}(\varepsilon,\vec p)$ is evident Lorentz invariant obeying the
following normalization:
\begin{equation}
\int S_{DS}(\varepsilon,\vec p)d\varepsilon d^3p\;=\;\int
S_{DS}(p)d^4p\;=\; A
\end{equation}
(here $p_0=m+\varepsilon$, so $dp_0=d\varepsilon$).\\

The calculations were performed for $^{12}C,\ ^{14}N,\ ^{27}Al,\
^{40}Ca,\ ^{56}Fe$ and $^{63}Cu$. The reason is as follows. As
mentioned above the necessary information for the calculations is that
about the proton and neutron density distributions.
 The former
 is available for throughout the whole periodic system\cite{8b}, but it
is not the case for the latter: the neutron densities are available
only for doubly closed-shell nuclei $^{16}O,\ ^{40}Ca,\ ^{90}Zr$ and
$^{208}Pb$\cite{9b}. That's why we confined ourselves by nuclei with a
small neutron excess: the density distributions per nucleon are nearly
the same for protons and neutrons in these nuclei\cite{10b}.\\

To calculate the eigen functions the Bonn B \cite{6b} and
OSBEP\cite{OSBEP} NN-potentials were used
\footnote{For the deuteron the Bonn B wave function was used in both
cases.}.  In both cases the results are very close to each other.
 The difference never exceeds 0.5\% for $x<0.6$ and is less then the
experimental error bars in the domain where the ratio (\ref{ratio})
$R_{th}>1$.

\section{Deep inelastic cross section\\ on nuclear target}
The DIS cross section is usually written in terms of the structure
function $F_2(x,Q^2)$, that is the cross section of electron-nucleon
interaction
\begin{equation}
\label{dis}
\frac{d\sigma}{dxdQ^2}\simeq\frac{4\pi\alpha^2}{xQ^4}
\left((1-y+\frac{y^2}2)F_2(x,Q^2)-\frac{y^2}2F_L(x,Q^2)\right)
\end{equation}
where we neglect the nucleon mass $m^2_N=m^2$ in comparison with the
total energy square $s=(k+p)^2 >> m^2$. Here: $k,q,p$ are the
4-momenta of the incoming electron, heavy photon and the target nucleon
repectively. $Q^2=-q^2$, $x=Q^2/2(p\cdot q)$ and $y=(q\cdot p)/(k\cdot
p)$. $\alpha=1/137$ is the electromagnetic coupling.\\

As a rule the data are taken at rather small $y$,
where the coefficient $y^2/2$ in front of the longitudinal part
($F_L$) is small.  Next, the ratio $R^L=F_L/F_2 \sim 0.2$ is not large.
Moreover, unlike the $F_2$, the function $R^L$ does not appear to depend
on atomic number $A$\cite{Arn}.

Therefore the ratio of cross sections is given usually in terms of
the ratio of structure functions $F_2$.\\

In order to compare our results with the data, where experimentalists
already accounted for the difference between proton and neutron,
 we write the structure function of nucleus as
$$\frac 1A
F_{2A}(x,Q^2)=\frac 1A\left( ZF_{2pA}+NF_{2nA}\right)=$$
\begin{equation}
\label{iso}
=\frac{F_{2nA}(x,Q^2)+F_{2pA}(x,Q^2)}2 +\frac{N-Z}{2A}(F_{2nA}-F_{2pA})
\end{equation}
and select the isospin I=0 part of $F_2$ given by the first term of
(\ref{iso}). The ratio which we will discuss reads
\begin{equation}
\label{ratio}
R_{kin}(x,Q^2)=\frac{F_{2nA}(x,Q^2)+F_{2pA}(x,Q^2)}{F_{2D}(x,Q^2)}\; .
\end{equation}

The structure function of the proton in nucleus
\begin{equation}
\label{f2p}
F_{2pA}(x,Q^2)=\frac 1Z\sum^{(p)}_\lambda
\nu_{p\lambda}F_{2p\lambda}(x,Q^2)\; ,
 \end{equation}
where $\nu_{p\lambda}$ is the actual number of protons on the level
$\lambda$ ($\nu_{p\lambda}=2j+1$ for the completely ocupied
shell). Note that in the experimental data the variable $x$ was
calculated assuming the proton momentum $p_N$ equal to the momentum of
a free proton at rest, $p_N=(m_N,0,0,0)$. However to single out the
precise "kinematics" one must account for the change of the nucleon
structure function (parton distibutions) in medium caused by the change
of the Bjorken variable $x=Q^2/2(p\cdot q)$.
In other words calculating the momentum fraction $x'$ carried by the
quark we need to use the precise four
momentum of the nucleon in medium.  That is
\begin{equation}
x'=\frac{Q^2}{2(pq)}\;=\;\frac{Q^2}{2(p_0q_0-\vec p\vec q)}\;=\;
\frac {mx}{m+\varepsilon_{\lambda}-\beta p t}\; ,
\label{xn}
\end{equation}
where $\beta=|\vec q|/q_0=(1+\frac{4m^2x^2}{Q^2})^{1/2}$
 and the variable $t$ is the cosine of the angle between $\vec p$
and $\vec q$.\\

Next we have to note that the structure function $F_2$, which
at the LO reads
\begin{equation}
F_2=\sum_f e^2_f(xq_f(x)+x\bar q_f(x))
\label{F2}
\end{equation}
( $e_f$ is the electric charge of the quark of flavour $f$)\\
contains two factors: the quark(antiquark) distribution $q(x)$($\bar
q(x)$) and the kinematical factor $x$. The origin of this kinematical
factor is as follows. The covariant quantity is not the cross section
but discontinuity of the dimensionless interaction amplitute
$ImA\simeq s\sigma$. Going from the amplitude $A$ to the cross section
$\sigma\propto 1/Q^2$ we obtain the factor $x_A=Q^2/2(pq)$ which
corresponds to the true nucleus target and must be calculated as
$x_A=AQ^2/2m_A\nu$, where $m_A$ is the mass of nucleus and $\nu=q_0$ is
the photon energy in the nucleus rest frame. Note that in the
final expressions (\ref{f2lam},\ref{f2D}) we use the value of
$m_A$ calculated within the doorway formalism,
that is $m_A=\sum_\lambda^{(p)}\nu_{p\lambda}(m+\varepsilon_\lambda)
+\sum_\lambda^{(n)}\nu_{n\lambda}(m+\varepsilon_\lambda)$,
 which is about 3\%
less than the true mass of a nucleus in the ground
state\footnote{For example, the 'doorway' mass of $^{40}Ca$ is
$m(doorway)=0.968m(ground\; state)$}.
 This is equivalent to the prescription given in \cite{FS},
where the authors accounted for the relativistic flux factor and used
the baryon charge conservation to normalize the spectral functions
 $S_p(\varepsilon,\vec p)$ and $S_n(\varepsilon,\vec p)$.  \\

Thus in (\ref{f2p}) we need to calculate the function

$$F_{2p\lambda}(x,Q^2)=\frac 12
\int^{p_{\lambda}}_0f_\lambda(p)p^2dp \int^1_{-1}
\frac{x_A}{x'}
F_{2p}(x',Q^2) dt\; +\;$$
\begin{equation}
\label{f2lam}
+\;\frac 12\int_{p_{\lambda}}^\infty f_\lambda(p)p^2dp
\int^{p_\lambda/p}_{-1}
\frac{x_A}{x'}
F_{2p}(x',Q^2) dt\; .
\end{equation}

Here: $p_\lambda=((1-x)m+\varepsilon_\lambda)/\beta$,
and  $f_\lambda(p)$
was defined in sect.2.2
\footnote{Strictly speaking (\ref{f2lam})
 is correct for a positive  $p_\lambda$ only. When $x$ is
close to 1 and $p_\lambda$ becomes negative one has to keep only the
last term in (\ref{f2lam}) with the integration from $-p_\lambda$ up
to $\infty$. In this case the values of $t < 0$ and $x'<x$. So the
quantity $F_{2\lambda}$ has non zero value even at $x=1$. Note however
that for experimentally available $x$ values the quantities
$p_\lambda$ never become negative.}\\

Exactly the same formulae is used for the neutron in nucleus.\\
For the deuteron
$$F_{2D}(x,Q^2)=\frac 12
\int^{p_D}_0f_D(p)p^2dp
\int^1_{-1}
\frac{x_D}{x'_D}
(F_{2p}(x'_D,Q^2)+F_{2n}(x'_D,Q^2))dt\; +$$
\begin{equation}
\label{f2D}
+\; \frac 12\int_{p_D}^\infty f_D(p)p^2dp
\int^{p_D/p}_{-1}
\frac{x_D}{x'_D}
(F_{2p}(x'_D,Q^2)+F_{2n}(x'_D,Q^2))dt
\end{equation}
with
$$x'_D=\frac {mx}{m_D-\sqrt{p^2+m^2} - \beta pt} $$
and
$p_D=(\beta(m_D-mx)-\sqrt{(m_D-mx)^2+(\beta^2-1)m^2}\ )/(\beta^2-1)$;
 $f_D(p)$ is just the sum of the squared monopole and quadrupole
components of the deuteron wave function;
$m_D$
is the deuteron mass.
Note that denominator in the expression for
$x'_D$ corresponds to the kinematics where the spectator nucleon is on
mass-shell.\\

The $F_{2p}(x,Q^2)$ and $F_{2n}(x,Q^2)$ free nucleon structure
functions were calculated using the MRST2002 NLO parametrization
\cite{MRST} obtained from the global parton analysis.

\section{Discussion}
The results of calculations are presented in Table 1-5 and
Fig.1.  The predictions made using the Bonn-B and OSBEP potentials
are very close to each other. So we present the results for the case
of Bonn B potential only.\\
\begin{table}[htb]
$$\begin{array}[t]{|c c| c c| c|}

\hline

 ^{12}C & &  NA-037 & NMC\cite{-3} &  \\
\hline
    x &    Q^2  &  R_{exp} & (\pm) &   R_{kin} \\
\hline
  .125 &   12.0 &   1.032 & ( .012) &   0.997 \\
  .175 &   15.0 &   1.011 & ( .015) &   0.994 \\
  .250 &   20.0 &   1.010 & ( .015) &   0.990 \\
  .350 &   27.0 &   0.971 & ( .020) &   0.985 \\
  .450 &   32.0 &   0.975 & ( .029) &   0.985 \\
  .550 &   37.0 &   0.925 & ( .043) &   0.999 \\
  .650 &   41.0 &   0.873 & ( .064) &   1.052 \\
\hline
\end{array}$$
\caption{The ratio of structure functions $F^A_2$ measured on carbon
 to that on deuteron. The values of $Q^2$ are given in GeV$^2$.}
\end{table}

\begin{table}[htb]
$$\begin{array}[t]{|c c| c c| c|}

\hline
 ^{14}N & &   NA-4 &  BCDMS\cite{-5} &       \\

\hline
    x &    Q^2  &  R_{exp} & (\pm) &   R_{kin} \\
\hline
  .100 &   32.0 &   1.018 & ( .039) &    0.997  \\
  .140 &   40.0 &   1.018 & ( .031) &    0.995  \\
  .180 &   49.0 &   1.002 & ( .024) &    0.993  \\
  .225 &   56.0 &   1.035 & ( .025) &    0.990  \\
  .275 &   56.0 &   1.024 & ( .027) &    0.988  \\
  .350 &   67.0 &   0.983 & ( .025) &    0.985  \\
  .450 &   77.0 &   0.941 & ( .031) &    0.985  \\
  .550 &   84.0 &   0.891 & ( .047) &    0.999  \\
  .650 &   96.0 &   0.826 & ( .075) &    1.053  \\

\hline
\end{array}$$
\caption{The ratio of structure functions $F^A_2$ measured on nitrogen
 to that on deuteron. The values of $Q^2$ are given in GeV$^2$.}
\end{table}

\begin{table}[htb]
$$\begin{array}[t]{|c c| c c| c|}

\hline

 ^{40}Ca  & &  NA-037 &   NMC\cite{-4} &   \\

\hline
    x &    Q^2  &  R_{exp} & (\pm) &   R_{kin} \\
\hline
  .113 &    4.3 &   0.994 & ( .010) &   0.998  \\
  .138 &    5.1 &   1.007 & ( .012) &   0.996  \\
  .175 &    6.2 &   1.001 & ( .011) &   0.994  \\
  .225 &    7.7 &   1.015 & ( .014) &   0.990  \\
  .275 &    9.1 &   0.998 & ( .018) &   0.986  \\
  .350 &   11.0 &   0.996 & ( .019) &   0.981  \\
  .450 &   14.0 &   1.024 & ( .031) &   0.978  \\
  .600 &   17.0 &   0.955 & ( .038) &   1.005  \\

\hline
\end{array}$$
\caption{The ratio of structure functions $F^A_2$ measured on calcium
 to that on deuteron. The values of $Q^2$ are given in GeV$^2$.}
\end{table}

\begin{table}[htb]
$$\begin{array}[t]{|c c| c c| c|}

\hline

 ^{56}Fe  &   &  NA-4 &  BCDMS\cite{-6} &  \\

\hline
    x &    Q^2  &  R_{exp} & (\pm) &   R_{kin} \\
\hline
  .100 &   22.0  &  1.057 & (.021) &   0.996   \\
  .140 &   25.0  &  1.046 & (.020) &   0.994   \\
  .180 &   29.0  &  1.050 & (.018) &   0.991   \\
  .225 &   46.0  &  1.027 & (.019) &   0.988   \\
  .275 &   49.0  &  1.000 & (.021) &   0.984   \\
  .350 &   59.0  &  0.959 & (.020) &   0.979   \\
  .450 &   72.0  &  0.923 & (.028) &   0.977   \\
  .550 &   72.0  &  0.917 & (.040) &   0.991   \\
  .650 &   72.0  &  0.813 & (.053) &   1.047   \\
\hline
\end{array}$$
\caption{The ratio of structure functions $F^A_2$ measured on iron
 to that on deuteron. The values of $Q^2$ are given in GeV$^2$.}
\end{table}

\begin{table}[htb]
$$\begin{array}[t]{|c c| c c| c|}

\hline

 ^{63}Cu &  &   NA-037 &  NMC\cite{-7} &  \\

\hline
    x &    Q^2  &  R_{exp} & (\pm) &   R_{kin} \\
\hline

  .123 &   11.0 &   1.041 & ( .026) &   0.996  \\
  .173 &   16.1 &   1.031 & ( .023) &   0.993  \\
  .243 &   19.3 &   1.018 & ( .024) &   0.988  \\
  .343 &   25.8 &   0.962 & ( .032) &   0.981  \\
  .444 &   36.0 &   0.959 & ( .047) &   0.978  \\
  .612 &   46.4 &   0.918 & ( .056) &   1.016  \\

\hline
\end{array}$$
\caption{The ratio of structure functions $F^A_2$ measured on copper
 to that on deuteron. The values of $Q^2$ are given in GeV$^2$.}
\end{table}

Recall that here we \underline{assume} the parton distributions
 inside the nucleon in nucleus to be the same as that for the free
nucleon and evaluate the pure kinematical effect of the boundness and
the motion of nucleon in nuclear matter.  Using the doorway states,
which are the correct eigen functions to describe the fast interaction
with one nucleon, we account for the full \underline{4-momemtum}
of the (target) nucleon and for the exitation of the ``residual''
nucleus $(A-1)$.

Thus the difference between the calculated value of $R_{kin}$ and the
data indicates the distortion of the parton wave function of a nucleon
placed in nuclear medium.\\

As expected the account of the boundness and Fermi motion of nucleons
in nuclei diminishes the cross section in the $x=0.2\ -\ 0.63$ interval.
Indeed, due to the boundness (and the fact that about 24 - 27 MeV is
spent for the exitation of the residual $(A-1)$ nucleus) the mean value
of shifted argument $x'$ (\ref{xn}) is larger than the value of
$x$ on a free nucleon. On the other hand in this domain the free
nucleon structure function $F_2$ falls down with $x$. Therefore we get
$R_{kin}<1$.\\

At a large $x$, close to 1, the details of angular integration (over
$t$ in (\ref{f2lam})) become important. For a negative $t$, due to a
Fermi motion, there is a region where $x'<x$ (see (\ref{xn})). Thanks
to the contribution coming from this region the value of $R_{kin}$
becomes larger than 1 for $x>0.65\ -\  0.7$.\\

Clearly, besides the Fermi motion there should be some dynamical
effects. At a large $x$ the growth of the ratio $R(x,Q^2)$  with $x$ is
usually atributed to a short range nucleon-nucleon
correlations\cite{FS2} or to a multiquark bags\cite{bag} (see for a
details the reviews \cite{Arn,FS2} and reference therein).  However,
contrary to the conventional expectations, the theoretical value of
 $R_{kin}$ resulting after account of the Fermi motion in the doorway
states is even \emph{larger} than the value $R_{exp}$ measured
experimentally\footnote{Since the same effect was observed both at
relatively low $Q^2$ in SLAC data and for a larger $Q^2$ at CERN this
can not be explained by the account of the mass correction.}.

This
means that in nuclear medium the (one nucleon) parton distribution
becomes softer, that is the probability to find a parton with $x>0.45$
inside the in-medium-nucleon is less than that in a free nucleon. In
other words in medium the quark distribution is shifted towards a
lower $x$, leading to the decrease of quark density at $x>0.45$ and a
larger quark density at a lower $x\sim 0.1\ -\  0.2$.\\

 Next at small $x<0.2$ the partons from different
(neighbouring) nucleons start to overlap and to interact with each other.
Indeed, according to uncertainty principle
 the characteristic size of localization is $\Delta r\sim 1/mx$ and for
$x<0.2$ the value of $\Delta r > 1$fm
 becomes comparable with the nucleon-nucleon separation.
 At a very low $x$ the partons screen each other and this shadowing
correction results in decreasing of $R(x,Q^2)$. Another way to describe
this effect is to say that two low-$x$ partons from two different
nucleons recombine into one parton. However the whole energy must be
conserved. This leads to the antishadowing (growth of the parton
density)\cite{NZ} (see the reviews\cite{Arn,FS2} for more details)
 just in the region ($x\sim 0.1\ -\ 0.2$) of the begining of
recombination.  On the other hand this antishadowing effect is expected
to reveal itself more in the gluon distributions than in the quark
structure function.

Thus it is not surprising that
 in the interval $0.2<x<0.45$ the ratio given by the pure kinematical
effects $R_{kin}$ (\ref{ratio})is close (within the error bars) to that
observed experimentally $R_{exp}$.\\

Note that, at large $x$, the Fermi motion is not negligible, even
for the  deuteron. The ratio $R_{D,kin}=F_{2D}/(F_{2p}+F_{2n})$
is close to one for $x<0.65$, but it noticeably differs from one
for $x>0.75$, reaching values of $R_{D,kin}=$1.07 (1.42) at
$x=$0.75 (0.85); see Table 6.

\begin{table}[htb]
$$\begin{array}[t]{|c|r r r r r r c|}
\hline
 x  & Q^2=\; 5 & 10 & 20 & 50 & 100 & 200 & GeV^2 \\
\hline

   .05 & R_{kin}=\;   1.000 & .999 & .999 & .999 &  .998 &  .998 & \\
    .10 & R_{kin}=\;\;\;  .999  & .999 & .998 &  .998 & .997 &  .997 &
  \\

    .14 & R_{kin}=\;\;\;  .998 &  .998 & .997 &  .997 & .996 &  .996 &
  \\

    .20 & R_{kin}=\;\;\;  .996 &  .996 & .995 &  .995 & .994 &  .994 &
  \\

   .35 & R_{kin}=\;\;\;  .990 &  .990 & .989 &  .989 & .989 &  .989 & \\

    .45 & R_{kin}=\;\;\;  .987 &  .987 & .987 & .987  & .987 &  .987 &
  \\

    .55 & R_{kin}=\;\;\;  .988 &  .988 & .988 & .989  & .990 & .991  &
  \\
   .65 & R_{kin}=\;  1.005 & 1.004 & 1.004 & 1.006 & 1.008 &  1.011
 & \\
   .75 & R_{kin}=\;  1.080 & 1.073 & 1.071 & 1.075 & 1.080 & 1.085 &
 \\
  .85 & R_{kin}=\;  1.440 & 1.396 & 1.382 & 1.391 & 1.408 & 1.431 &
  \\

\hline

\end{array}$$
\caption{The kinematical part $R_{kin}$ of the ratio
$F^d_2/[F^p_2+F^n_2]$ calculated using the Bonn-B potential}
\end{table}

 An analysis performed by the MRST group
shows that if this effect is included then one obtains practically the
same partons, but the description of the high $x$ deuteron data is much
improved; with $\chi^2$ reduced by 20 for the 12 deuteron data points that are fitted at
$x=0.75$ \footnote{We thank R.S.Thorne and A.D.Martin for
discussions and for performing a new analysis using our Fermi motion in
the deuteron.}.\\

\begin{figure}
\centerline{\vspace{0.2cm}\hspace{0.1cm}
\epsfig{file=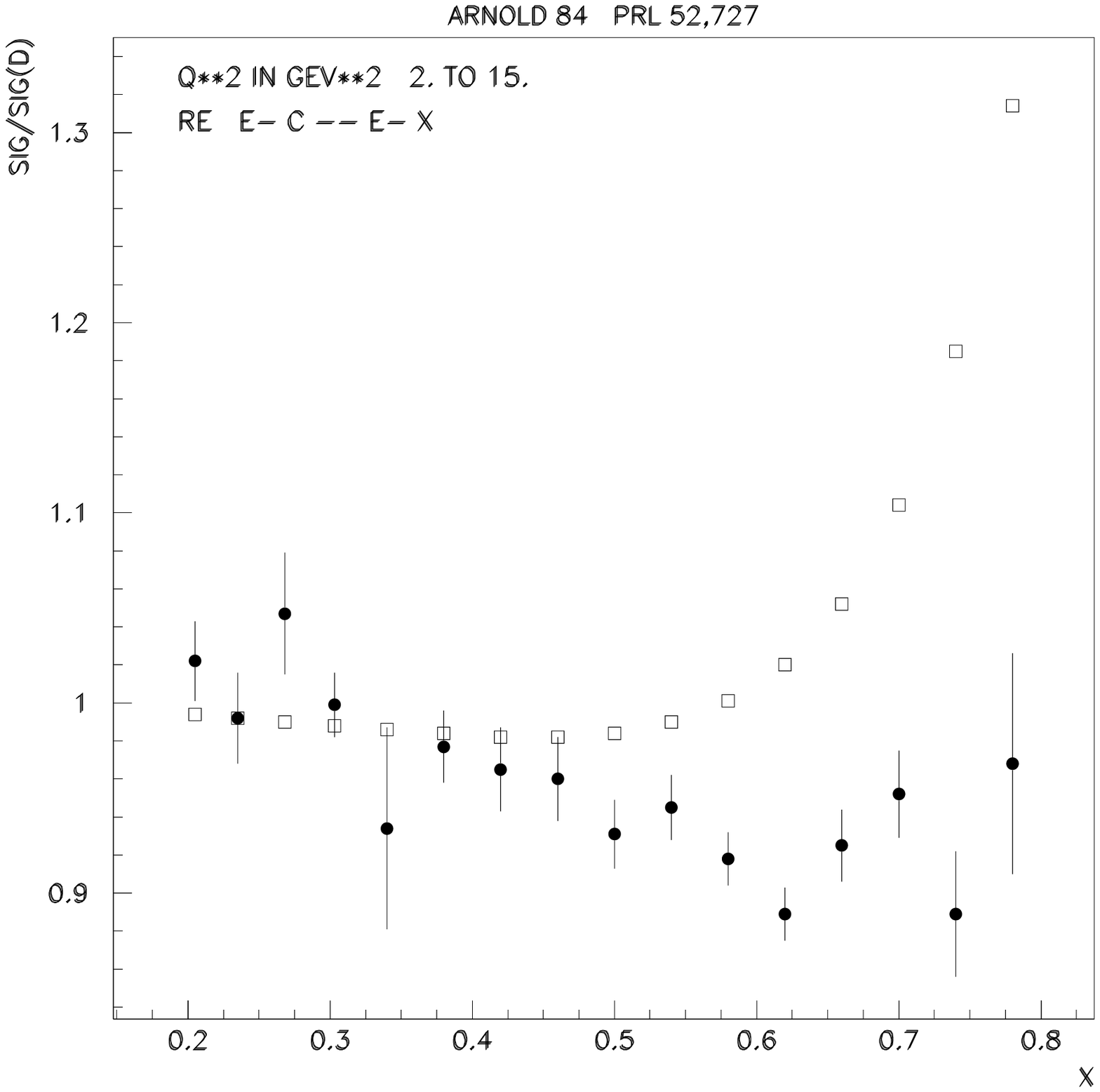,width=7cm}
\epsfig{file=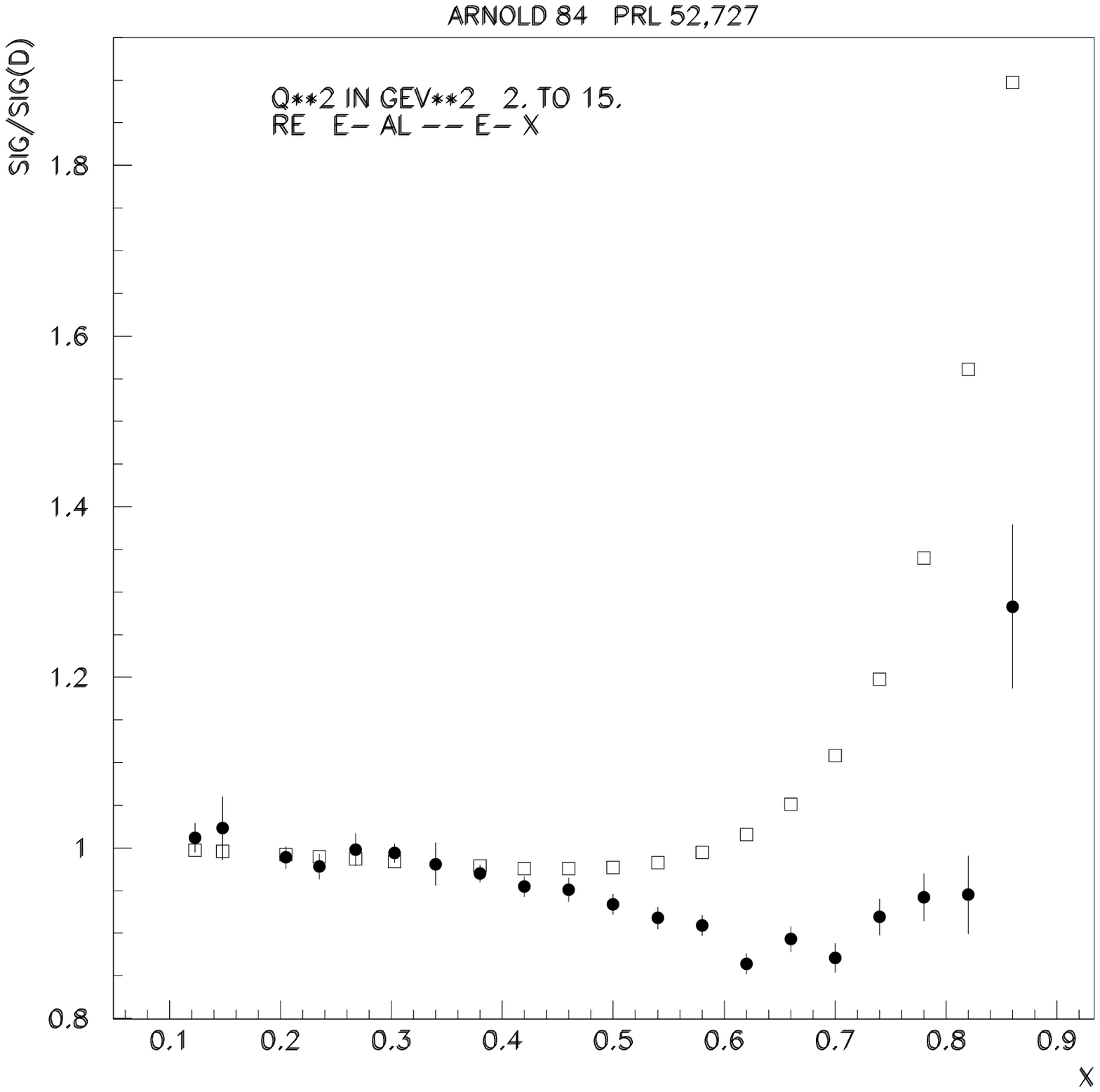,width=7cm}
}
\vspace*{-0.5cm}
\centerline{\vspace{0.2cm}\hspace{0.1cm}
\epsfig{file=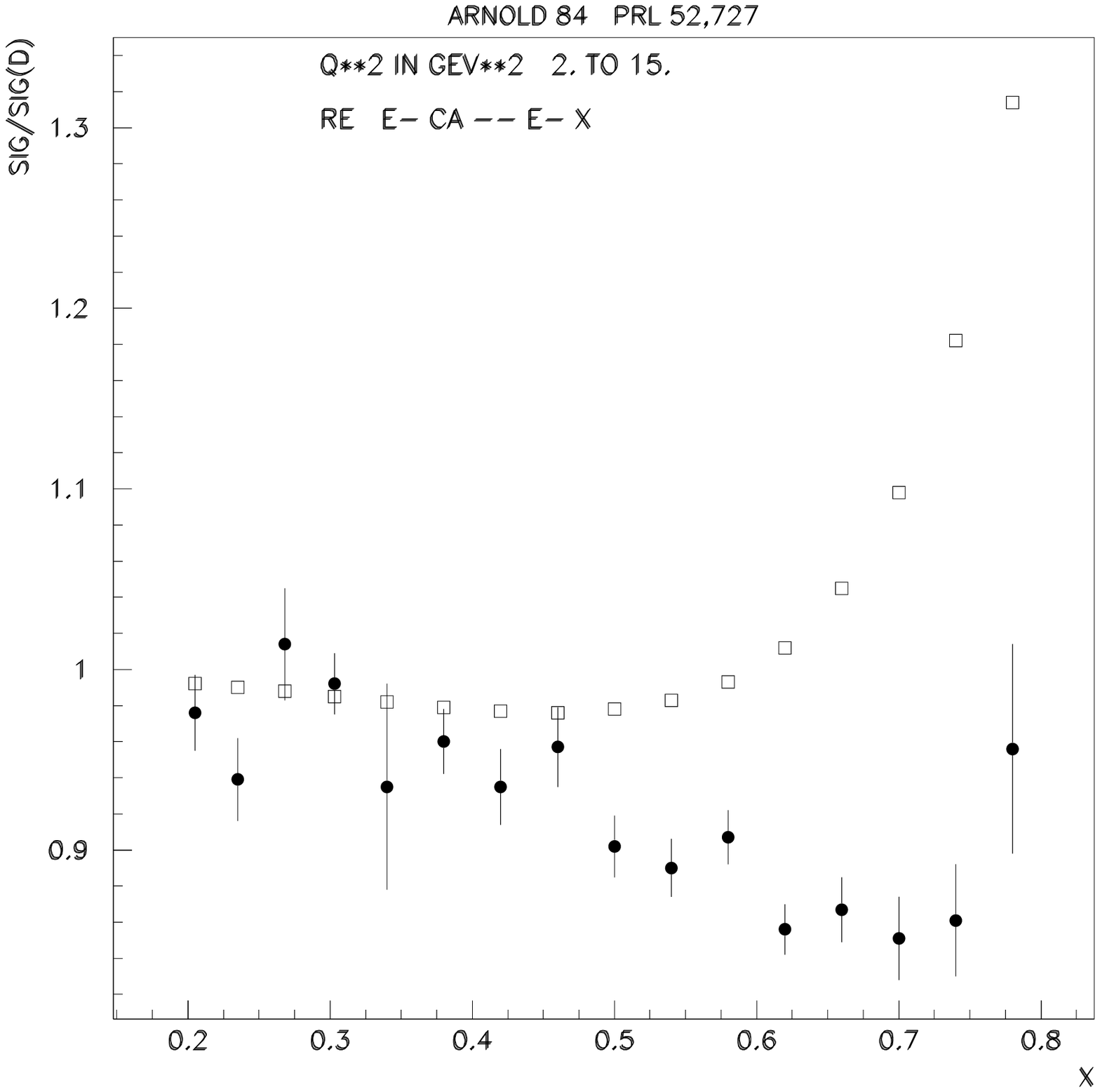,width=7cm}
\epsfig{file=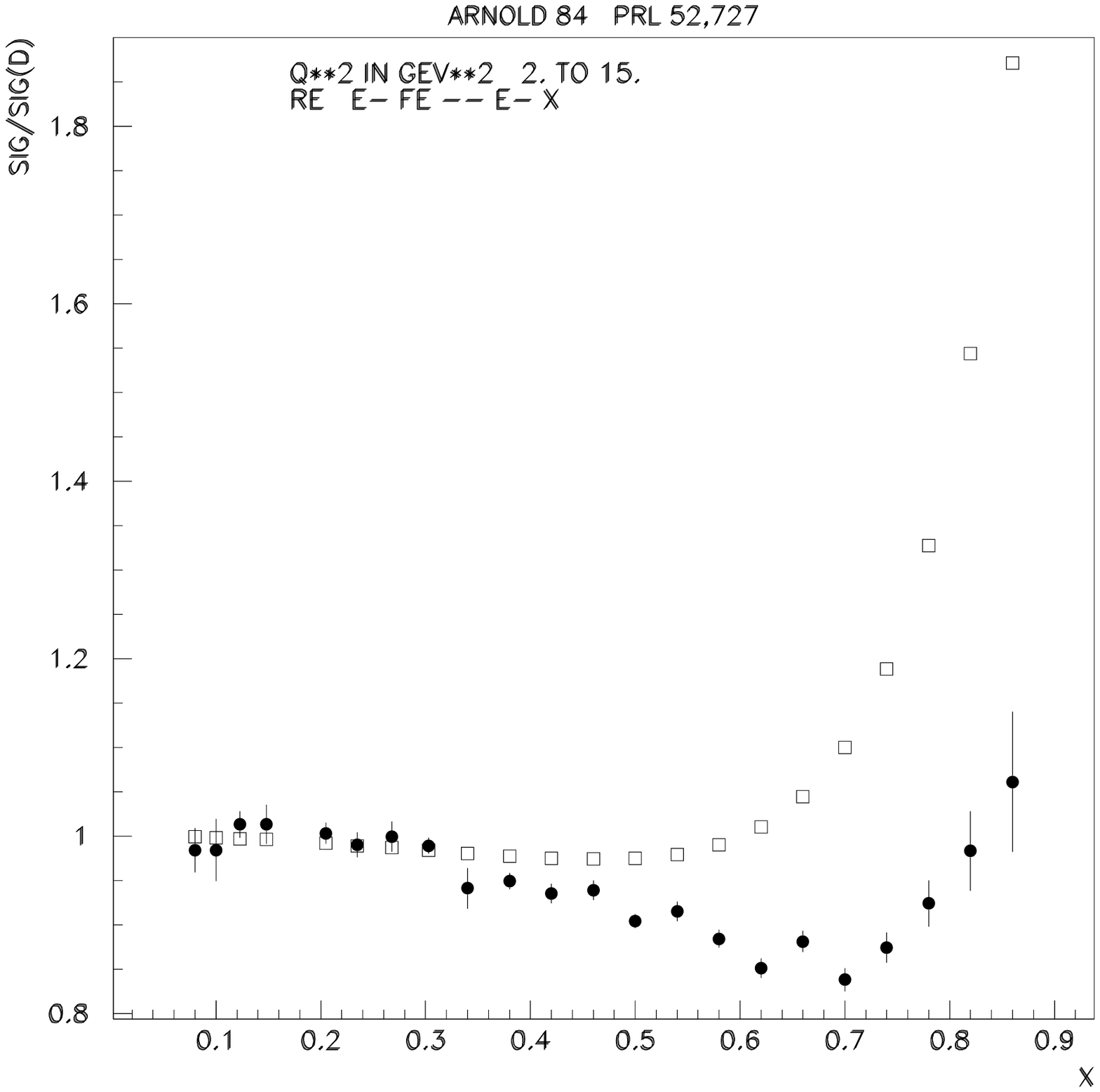,width=7cm}
}
\caption{ Fig.1 The ratio of structure function $F^A_2$
measured on nucleus $A$ to that on deuteron. $Q^2=5$ GeV$^2$.
 The data are taken from \cite{SLAC}. The empty bar
is the ratio $R_{kin}$ calculated using the Bonn-B potential.}
\end{figure}

After the present work was completed we have read the recent paper of
A.Molochkov \cite{Mol} where a little bit another (but not quite
different from that used here) prescripton was proposed to account for
the boundness and momentum distribution of nucleons. A.Molochkov had
considered the ratio of the $^4$He to deuteron structure functions.
The shortness of his prescription is the assumption that both the
nucleon structure function $F_2$ and the momentum distribution of the
nucleons in nucleus $f^N(P_A,p)$ are regular (i.e.  have no
singularities) with respect to $p_0$. Besides this some terms,
coming from the differentiation of the nucleus
$(A-1)$ propagator and the factor $1/(p_0+\sqrt{m^2+p^2})^2$
(corresponding to the antinucleon pole) in the nucleon propagator,
which are proportional to the binding (or nuclear exitation) energy,
were omitted in \cite{Mol}.  We hope that our approach, based
on the 'doorway' formalism is more precise. Moreover, in terms of
the Molochkov's integral our result may be
obtained by closing the integration contour over $p_0$  in the upper
half-plane (on the pole corresponding to the residue $(A-1)$ nucleus)
instead of the lower one as it was done in\cite{Mol}.


However we are planning to
compare both approaches in the forthcoming paper, using the doorway
eigen functions to describe the distributions of nucleons in a heavier
nuclei.


\end{document}